\documentclass[12pt]{article}
\usepackage{graphics,psfrag,epsfig}
\usepackage{psfig}
\usepackage{epsf}
\usepackage{float}
\usepackage{latexsym}
\usepackage{rotating}
\usepackage[german,USenglish]{babel}
\newcommand{\beq}{\begin{equation}}
\newcommand{\eeq}{\end{equation}}
\newcommand{\beqa}{\begin{eqnarray}}
\newcommand{\eeqa}{\end{eqnarray}}

\newcommand{\bma}{\begin{array}{cc}}
\newcommand{\ema}{\end{array}}

\def\3{{\ss}}

\def\vek #1 {\overrightarrow {#1}}

\begin{document}


\begin{center}
{{\large\bf Indications for the Nonexistence of Three-Neutron Resonances near
    the Physical Region}}
\end{center}


\begin{center}

A.\ Hemmdan$^{\dagger, \, \star, }$ \footnote{email: Amel.Hemdan@tp2.ruhr-uni-bochum.de},  
W.\ Gl\"ockle $^{\dagger , \,}$\footnote{email:
  Walter.Gloeckle@tp2.ruhr-uni-bochum.de}, and H.\
Kamada$^{\ddagger,}$\footnote{email: kamada@mns.kyutech.ac.jp} 
\bigskip

$^\dagger${\it Institut f\"ur Theoretische Physik II, Ruhr-Universit\"at
  Bochum, D-44870 Bochum, Germany}\\ 

$^\star${\it Department of Physics, Faculty of Science, South Valley University, Aswan, Egypt}\\

$^\ddagger${\it Department of Physics, Faculty of Engineering, Kyushu Institute
   of Technology, \\ 1-1 Sensuicho, Tobata, Kitakyushu 804-8550, Japan}

\end{center}

\vspace{.3in}

\thispagestyle{empty} 

\date{\today}

\begin{abstract}

The pending question of the existence of three-neutron resonances near the
physical energy region is reconsidered. Finite rank neutron-neutron forces are used
in Faddeev equations, which are analytically continued into the unphysical
energy sheet below the positive real energy axis. The trajectories of the
three-neutron $S$-matrix poles in the  states of total angular momenta and
parity $J^\pi={\frac{1}{2}}^\pm$  and $J^\pi={\frac{3}{2}}^\pm$ are traced out
as a function of 
artificial enhancement factors of the neutron-neutron  forces. The final
positions of the $S$-matrix poles  removing the artificial factors  are
found in all cases to be far away from the positive real energy axis, which
provides a strong indication  for the nonexistence of nearby three-neutron
resonances. The pole trajectories close to the threshold $E=0$ are also
predicted out of auxiliary generated three-neutron bound state energies
using the Pad\'e method and agree very well  with the directly
calculated ones. 
\end{abstract}
PACS numbers: 21.45.+v,24.30.Gd,25.70.Ef

\newpage

\section*{I Introduction}

Controversial experimental hints on the existence of three-neutron ($3n$)
resonances [1-4] and more recently even on four-neutron resonances [5]
occurred again and again. No definite experimental conclusion on $3n$
resonances seem to exist. On the theoretical side we are aware of various
attempts to investigate states of $3n$'s but also here the situation is pretty
unsettled. Obviously the optimal theoretical present day insight would be if
the most modern neutron-neutron ($nn$) forces from the set of high precision
 $NN$ potentials together with $3n$ forces of the $2\pi$-exchange type
 (Tucson-Melbourne [6] or Urbana type [7]) would be used and the energy
 eigenvalues of the $3n$ Schr\"odinger equation according to resonance
 boundary conditions would be determined in the unphysical energy sheet
 adjacent to the positive real energy axis. Unfortunately we can  not provide
 such an insight right now but think we can improve at least  on the  existing
 studies we are aware of. 

One of the possibly first studies on that list is a solution of the $3n$
 Faddeev equation based on a Yamaguchi rank 1 $nn$ force in the state $^1S_0$
 ~[8]. Only the $3n$ states of total angular momenta $J^{\pi}=1/2^-$ and
 $3/2^-$ (degenerate in this case) have been studied. The $nn$ force was
 artificially enhanced such that two and three neutrons were bound. Then the
 enhancement was reduced which has the consequence that both bound state
 energies move towards  $E=0$. For the forces used it happened that the $3n$
 binding energy moves faster than the $2n$ binding energy and thus hits the
 dineutron-neutron threshold to the left of $E=0$. The trajectory of the $3n$
 bound state energy continues  then into the unphysical sheet below the
 dineutron-neutron cut. In this manner a $3n$ resonance occurs, below the $3n$
 break-up threshold at $E=0$. Further decreasing the enhancement factor the
 $3n$ resonance trajectory moves up again exactly towards $E=0$ where it meets
 with the $2n$ bound state energy. Both energies coincide exactly at $E=0$
 corresponding still to an enhancement factor larger than 1. In further
 decreasing the potential strength till one reaches the physical value 1 the
 vanishing $2n$ bound state energy turns into the well established $2n$
 virtual state energy in the second sheet on the negative real energy axis and
 the $3n$ resonance disappears in a unphysical sheet which is different from
 the sheet adjacent to the positive real axis. In other words the complex
 resonance energy of that $3n$ resonance for $J^{\pi}=1/2^-$ and $3/2^-$ has
 no positive real part and therefore can not be felt for positive
 three-neutron energies. We have to emphasise that this refers to a $nn$ force
 acting only in the state $^1S_0$ and it will change if additional $nn$ force
 components will be added as will be shown in this paper.

A very much related study has been performed in [9] with essentially the same
result as in [8]. Using just bound state techniques [10] the necessary
enhancement factors on the $nn$ forces in $S$- and $P$-waves have been
determined to bind 3 neutrons near zero energy. Based on the Reid potential
[11] enhancement factors of the order of 4 have been found which make low
lying $3n$ resonances quite unlikely.

Another theoretical investigation [12] we are aware of for the state
  $J^{\pi}=1/2^-$ has been performed on the basis of a hyperspherical harmonic
  expansion. Local forces like the ones of Mafliet Tjon have been used and
  this only in the $nn$ state $^1S_0$. The expansion was truncated and the
  zeros of the Jost function were determined. It resulted a resonance energy
  around $E=-4.9-i~6.9$ MeV in an unphysical sheet. The authors found a
  strong sensitivity to the choice of the potential.

A further investigation [13] relied on a variational ansatz and used the
 complex scaling method which allows to treat a resonance problem like a bound
 state one. The Minnesota effective $nn$ force [14] together with tensor
 forces [15] have been applied. The authors find as the only candidate for a
 $3n$ resonance the state $J^{\pi}= 3/2^{+}$ and a prediction for its energy
 of $E=14-i~13$ MeV. 

Finally in a more recent work [16] the Faddeev equation in configuration space
 has been solved for the states $J^{\pi}=1/2^-$ and $3/2^{\pm}$ using the more
 realistic forces by Gogny {\it et al.} [17] and the Reid93 [18] potential in
 the $2n$ states $^1S_0$, $^3P_0$ and $^3P_2$ - $^3F_2$. Choosing  proper
 boundary conditions according to $3n$ resonances the complex energy
 eigenvalues are determined starting again from artificially enhanced forces
 and reducing their strengths gradually. Unfortunately with increasing
 negative imaginary parts of the complex resonance energies numerical
 instabilities occurred  and the trajectories could not be followed up until
 the physical values for the enhancement factors have been reached.

Now we would like to improve on all that by solving the Faddeev equations with
low rank $nn$ forces in all relevant partial wave states exactly and
determining the final resonance positions for the actual force strength as
prescribed by $nn$ phase shifts (assumed to be identical to the strong $pp$
phase shifts). In the light of our results we shall also comment on the
previous findings.

This paper is organised as follows. In section II we briefly review the set of
coupled Faddeev eigenvalue equations for a $3n$ system based on finite rank
forces and mention the steps needed for an analytical continuation into the
unphysical energy sheet adjacent to the positive real axis. The dynamical $nn$
force input and the resulting $3n$ resonance trajectories for the states
$J^{\pi}=1/2^{\pm}$ and $3/2^{\pm}$ are presented in section III. In section
IV we show how the part of the $3n$ resonance trajectory close to the
threshold $E=0$ can be predicted with the help of Pad\'e approximant's
 from a set
of $3n$ bound state energies. This serves as a test for our
results obtained through the analytical continuation of the Faddeev
equation. Though the final $3n$ resonance positions could not be reached with
that method, we think this illustration should be of interest since it works
beautifully for resonances close to the first threshold and requires only
bound state techniques. We summarise in section V.

\section*{II Formalism}

The Faddeev equation for three identical nucleons has the well known
form [19]
\begin{equation} 
\psi =G_{0}tP\psi\, ,
\end{equation}
where $G_{0}$ is the free $3N$ propagator, $t$ the $NN$ $t$-matrix and $P$ the
sum of a cyclical and anti-cyclical permutation of 3 objects. We regard that
equation in a momentum space representation and partial wave decomposed. It
results a set of coupled equations in two variables, p and q, which are the magnitudes
of two relative momenta. In the present investigation we shall use
$t$-matrices of finite rank. The set of two-dimensional equations changes then
into a set of coupled one-dimensional ones. They have the well known form 
\begin{eqnarray}
\theta _{\alpha }^{j}(q)=\int ^{\infty }_{0}dq\prime q\prime ^{2}\sum _{\alpha
  \prime }\sum _{km} \int ^{1}_{-1}dx\frac{g^{{\tilde{\alpha }}}_{j}(\pi _{1})}{\pi ^{l}_{1}}G_{\alpha \alpha \prime }(q,q\prime,x) &  & \nonumber \\
\frac{1}{(E-\frac{q ^{2}}{M}-\frac{q\prime ^{2}}{M}-\frac{q q\prime x}{M})}
 \frac{g^{\tilde{\alpha }^{\prime }}_{k}(\pi _{2})}{\pi ^{l\prime }_{2}}
\tau ^{{\tilde{\alpha }}^{\prime }}_{km}(z\prime )\theta ^{m}_{\alpha \prime }
(q\prime ) &  & 
\end{eqnarray}
and $\theta$ is connected to $\psi$ via 
\begin{equation}
\psi _\alpha (p , q) = { 1 \over { E - { p^2 \over M } - { 3 q ^2 \over {4 M }}  }}
\sum _{j,k} g^{\tilde \alpha} _ j (p) 
\tau _{j,k} ^{\tilde \alpha} (z) \theta_\alpha ^k (q). 
\end{equation}
The underlying two-body $t$-matrices are 
\begin{equation}
t(p,p\prime ,z)=\sum _{i,j}\, g_{i}(p)\, \tau _{i,j}(z)\, g_{j}(p\prime )
\end{equation}

The corresponding form factors $g(p)$ depend on the obvious two-body 
quantum numbers ${\tilde{\alpha}}\equiv (lsj)$. The two-body subsystem energy
$z=E-\frac{3q^{2}}{4M}$ is given in terms of the total energy $E$ and the
kinetic energy of relative motion of the third particle. Further $\alpha$
stands for the string of three-body quantum numbers 
\begin{equation}
\alpha \equiv (ls)j(\lambda \frac{1}{2})I(jI)J(t\frac{1}{2})T
\end{equation}
For all those by now standard notations we refer to [19]. Finally
$G_{\alpha\alpha\prime}(q,q\prime,x)$ are geometrical coefficients arising
from the partial wave representation of the permutation operator $P$ and
$\pi_{1}$ and $\pi_{2}$ are shifted arguments related to $P$ and depend on
$q$, $q\prime$ and $x$, see [19] for details.

We are interested in $3n$ bound state and resonance energies, the latter ones
emerging from bound state energies when the artificially increased interaction
strength is reduced. Since two neutrons are not bound in nature the only
unphysical sheet of interest is the one which can be reached through the cut
along $0\leq E<\infty$. It will be denoted in the following simply as the
unphysical sheet. While for  $E<0$  there is no singular denominator, for
$E>0$ the free propagator in Eq.\ (2) is not defined. The well established
manner to analytically continue the coupled set of equations in the energy 
into the unphysical sheet is to deform the path of integration into the 
lower half plane. This has been described for instance in [8] and [20].  
The singularities which are avoided in this manner arise from the free
propagator and the form factors. The $\tau$-function is not singular for
energies $z=E-\frac{3q^{2}}{4m}$ in the unphysical sheet under
investigation. A typical path in the complex $q$-plane (for both $q$ and
$q\prime$) together with domains of singularity arising from the free
propagator for the example $E=10.0-i~10.0$ MeV is shown in Fig.\ 1.
\begin{figure}
\centering
\psfig{file=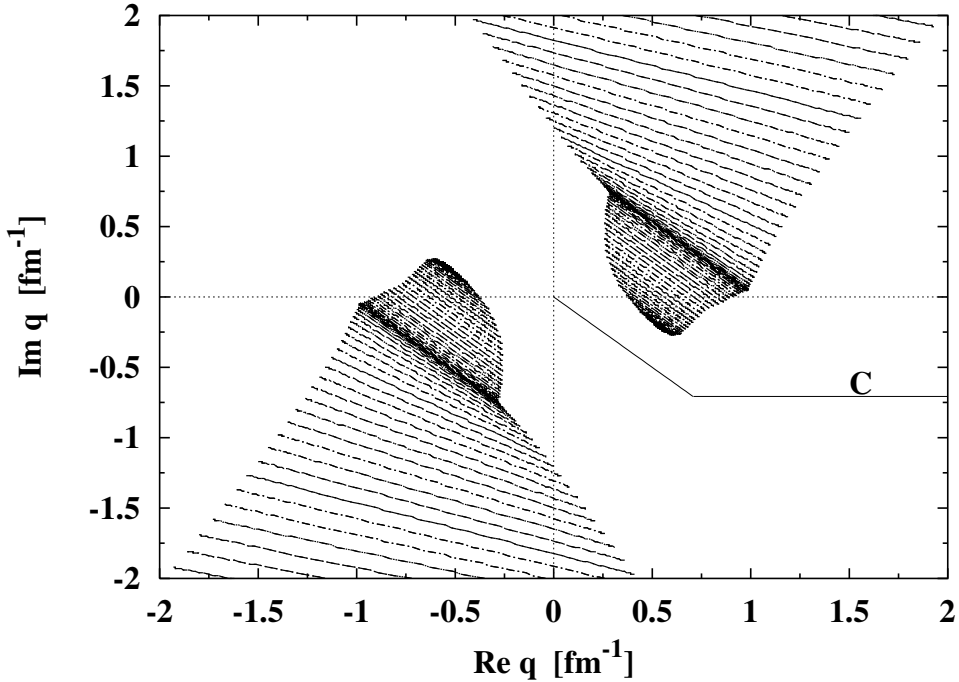,width=13cm}
\caption{The regions of singularities  from the free propagator for  the
complex energy $E=10.0-i~10$ MeV and a path of integration C.} 
\end{figure}

The strategy now  as pointed out in the introduction is to first enhance the
 $nn$ forces such that a $3n$ bound state exists. At the same time also
 two-neutron bound states might exist. Then by suitably chosen enhancement
 factors one can achieve that the two-body bound state energies coincide and
 move together towards $E=0$, when the  enhancements are reduced. In all cases
 we studied the $3n$ bound state  energies stayed always to the left of the
 two-body bound state energies until they reached $E=0$. This was not the case
 in [8], where for the degenerate states  $1/2^-$ and $3/2^-$ only the $^1S_0$
 force had been kept. Here we include on top $P$- and $D$-wave forces and such
 a coincidence of $3n$ and $2n$ bound state energies does not occur. Once the
 two-neutron bound state energies have disappeared we still have a  $3n$ bound
 state. Then reducing the enhancement factors further the $3n$ bound state
 energy will approach $E=0$ and then the energy eigenvalue $E$ will dive into
 the lower half plane of the unphysical sheet. There we shall follow its
 trajectory until the physical strengths of the $nn$ forces are reached. This
 defines the final positions of the $3n$ resonances. It is well known and
 easily seen that  what we call $3n$ resonance energies are poles of the $3n$
 $S$-matrix.

\section*{III Results}

We use $nn$ forces of rank 2. They are given in Appendix A. 
They describe $pp$ phase shifts (without Coulomb force) reasonably
well as documented in the Appendix.
We restrict our study to the states $^1S_0$, $^3P_0$,$^3P_2$ and  $^1D_2$. The
coupling between $^3P_2$ and $^3F_2$ is neglected. The $nn$ force in the state
$^3P_1$ is repulsive and therefore we did not include it in the main
investigation, since it has to be expected, that it will move the $3n$
resonance position further away from the positive real energy axis. At the end
we shall consider its effect and show that its influence has indeed repulsive
character but only a marginal  one and could therefore indeed be neglected.

We shall investigate now the 4 possible
states  $J^{\pi}= 1/2^{\pm}$ and $J^{\pi}= 3/2^{\pm}$ in turn.

\subsection*{3.1 The state $3/2^-$}

We use up to 10 three-neutron channels which are displayed in Table 1. To reach a $3n$
bound state it is sufficient to enhance just the $^3P_2$ $nn$ force and to
keep all the other force components at their physical values. As an example
for an enhancement factor we quote $\lambda_{^3P_2}=3.5$ where the 3 neutrons
in a 10 channel calculation are bound at $-7.79$ MeV. The threshold energy
$E=0$ is reached for $\lambda_{^3P_2}=3.23$. For the sake of future
comparisons we provide a few intermediate $3n$ resonance positions in Table
2. The $3n$ resonance trajectory is shown in Fig.\ 2 for a 10 channel
calculation. The trajectories for a smaller number of channels are very
similar.  The final resonance positions for an increasing number of channels in
 the order given in Table 1 are displayed in Table 3. Adding more $nn$ force
 components will certainly not change the final position in a significant
 manner in the sense that the position would come somewhere close to the
 positive real energy axis. If one adds the $^3P_1$ $nn$ force, which is of
 repulsive nature, we have 13 channels and  the final resonance position shifts
 by $-1.79-i~2.87$ MeV, which is marginal.

In Ref.[16] which can be considered to be the most realistic approach
towards three-neutron resonances carried out so far the rather realistic $nn$
forces by Gogny [17] and Reid93 [18] have been used. In contrast to our
simplified forces they also include the tensor force induced coupling
$^3P_2-^3F_2$. For the case of the Gogny potential the enhancement factor for
the  $^3P_2-^3F_2$  $nn$  force component was about 4.4 to have $3n$'s bound
at zero energy, while for our simplified force this was the smaller value of
$3.5$. In [16] reducing $\lambda_{^3P_2-^3F_2}$ to $3.7$ yields a resonance
energy of $4.95-i~3.75$ MeV, while at this value our $3n$ system is still
bound. Thus one might conjecture that in the more realistic case the $3n$
resonance would be even further shifted away from the positive real energy
axis. On the  other hand  the motion of our resonance energy for a change of $
\Delta\lambda=4.4-3.7=0.7$ from the threshold $E=0$ (corresponding to an
enhancement factor $2.8$) is  $5.58-i~4.04$ MeV, which is slightly larger.
Therefore a direct quantitative  comparison is not possible, though
qualitatively the results are similar. This is also true for the  case of
Reid93, also considered in [16], where  the enhancement factor $\lambda_{^3P_2-^3F_2}=3.25$  yields a resonance position of $5.30-i~3.53$ MeV.
\begin{table}
\begin{center}
\caption{\small The partial wave quantum number for the three neutron state
  $J^{\pi} = \frac{3}{2}^{-}$.}
\vspace*{0.3cm}
\begin{tabular}{||c|c|c|c|c|c||}
\hline\hline
$nn$ {\rm state} & $l$ & $s$ & $j$ & $\lambda$ & $I$   \\
[1.1ex]
\hline\hline 
 $^1S_0$  & 0 & 0 & 0 & 1 & 3/2      \\
 $^3P_2$  & 1 & 1 & 2 & 0 & 1/2     \\
 $^3P_0$  & 1 & 1 & 0 & 2 & 3/2     \\
 $^3P_2$  & 1 & 1 & 2 & 2 & 3/2     \\
 $^3P_2$  & 1 & 1 & 2 & 2 & 5/2    \\
 $^1D_2$  & 2 & 0 & 2 & 1 & 1/2     \\
 $^1D_2$  & 2 & 0 & 2 & 1 & 3/2     \\
 $^1D_2$  & 2 & 0 & 2 & 3 & 5/2     \\
 $^3P_2$  & 1 & 1 & 2 & 4 & 7/2    \\
 $^1D_2$  & 2 & 0 & 2 & 3 & 7/2      \\
[1.1ex]
\hline\hline
\end{tabular}
\label{table1}
\end{center}
\end{table}

\begin{table}
\begin{center}
\caption{\small Intermediate $3n$ resonance positions  together with the
enhancement factor $\lambda_{^3P_2}$ for a $10$ channel calculation and
$J^{\pi} = \frac{3}{2}^{-}$.} 
\vspace*{0.3cm}
\begin{tabular}{||c|c|c||}
\hline\hline
{$\lambda _{^3P_{2}}$}  & $E^{res.}_{3n}$ [MeV]  \\
[1.1ex]
\hline\hline 
 3.22  & 0.20 -$i$ 0.003    \\
 3.20  & 0.57 -$i$ 0.03     \\
 3.00  & 3.51 - $i$ 1.45      \\
 2.00  & 5.40 -$i$ 19.17     \\
 1.50  &-1.95 -$i$ 28.87     \\
 1.00  &-12.13-$i$ 37.96    \\
[1.1ex]
\hline\hline
\end{tabular}
\label{table2}
\end{center}
\end{table}

\begin{figure}
\centering
\psfig{file=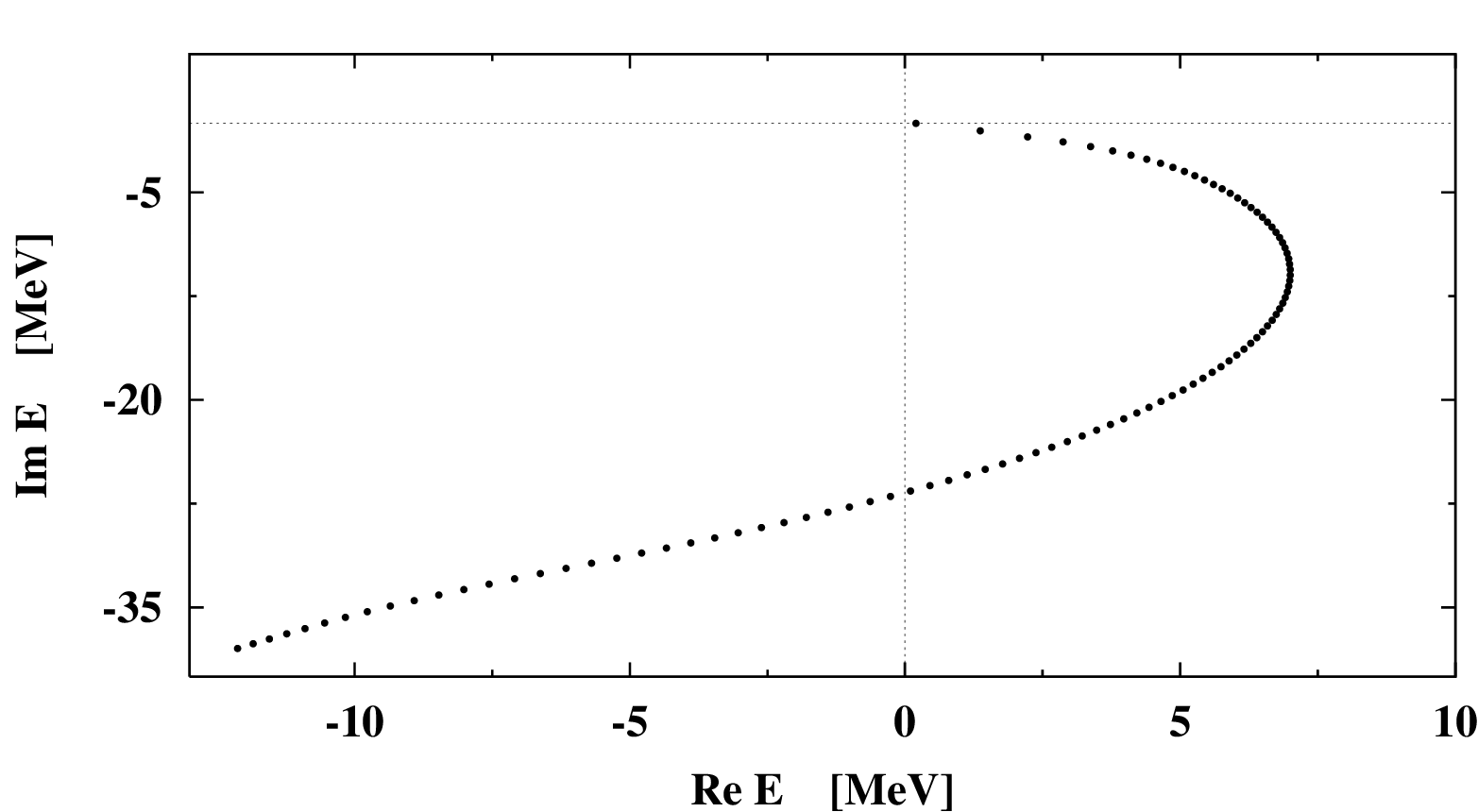,width=13cm}
\caption [Fig.3] {The resonance pole trajectory for the state 3/2$^-$.} 
\end{figure}

\begin{table}
\begin{center}
\caption{\small The final $3n$ resonance  positions for the state
$J^{\pi} = \frac{3}{2}^{-}$ including different number of channels.}

\vspace*{0.3cm}
\begin{tabular}{||c|c||}
\hline\hline
{\rm Number of channels}  & {\rm Final resonance positions}  \\
[1.1ex]
\hline\hline 
 2  & -12.81-$i$ 33.22    \\
 3  & -13.14-$i$ 33.21    \\
 4  & -13.11-$i$ 33.17     \\
 5  & -13.33-$i$ 32.91     \\
 6  & -14.28-$i$ 33.37     \\
 7  & -12.43-$i$ 37.65    \\
 8  & -12.31-$i$ 37.74     \\
 9  & -12.31-$i$ 37.74    \\
 10 & -12.13-$i$ 37.96    \\
[1.1ex]
\hline\hline
\end{tabular}
\label{table3}
\end{center}
\end{table}
\vspace{2.5cm}

\subsection*{3.2 The state $3/2^+$}

Again we use up to $10$ channels, which are displayed in Table 4.  In order to
have a $3n$ bound state we had to enhance the $^3P_0$ and $^3P_2$ $nn$ forces
while the $^1S_0$ and $^1D_2$ forces could be kept at their physical
values. For the enhancement factors  $\lambda_{^3P_2}=3.5$ and
$\lambda_{^3P_0}=5.0$ $2n$ bound states for $^3P_2$ and $^3P_0$ just disappear
and the $3$ neutrons are bound at $-6.55$ MeV. Then we reduce both enhancement
factors. The way this is done is in principle arbitrary. Again for future work
we provide some intermediate $3n$ resonance positions along the trajectory in
the unphysical sheet in Table 5. The corresponding trajectory is shown in
Fig.\ 3. The dependence of the final position on the number of channels is
displayed in Table 6. Like for $3/2^-$ the final $S$-matrix pole position is
far away from the positive real energy axis. Again the effect of  $^3P_1$ $nn$
force is marginal (a shift of $-0.80-i~0.42$ MeV).

Now we compare to the work in Ref. [16].
The enhancement factors for the Gogny potential for the $^3P_0$ and
$^3P_2-^3F_2$ $nn$ force components are about $7.7$ and $4.9$ to generate $nn$
bound states at zero energy.  This is larger in both cases than for our
simplified potential. Correspondingly according to Ref.\ [16] for  $\lambda
_{^3P_0}=5.0$ and $\lambda_{^3P_2-^3F_2}=4.30$ one ends up with a resonance at
$5.74 - i~1.53$ MeV, while for such enhancement factors our $nn$ forces still
bind the $3n$ system. Thus also here one is tempted to conjecture that the
more realistic Gogny potential would lead to a final resonance position, which
is farther away from the real axis than what we find. For the Reid93 potential
the corresponding values are $5.93-i~1.55$ MeV for $\lambda _{^3P_0}=4.6$ and
$\lambda _{^3P_2-^3F_2}=3.50$.
\begin{table}
\begin{center}
\caption{\small The partial wave quantum number for the three neutron state
  $J^{\pi} = \frac{3}{2}^{+}$.}
\vspace*{0.3cm}
\begin{tabular}{||c|c|c|c|c|c||}
\hline\hline
 $nn$ {\rm state} & $l$ & $s$ & $j$ & $\lambda$ & $I$   \\
[1.1ex]
\hline\hline 

 $^1S_0$  & 0 & 0 & 0 & 2 & 3/2     \\
 $^3P_0$  & 1 & 1 & 0 & 1 & 3/2     \\
 $^3P_2$  & 1 & 1 & 2 & 1 & 1/2     \\
 $^3P_2$  & 1 & 1 & 2 & 1 & 3/2     \\
 $^1D_2$  & 2 & 0 & 2 & 0 & 1/2     \\
 $^3P_2$  & 1 & 1 & 2 & 3 & 5/2     \\
 $^3P_2$  & 1 & 1 & 2 & 3 & 7/2    \\
 $^1D_2$  & 2 & 0 & 2 & 2 & 3/2    \\
 $^1D_2$  & 2 & 0 & 2 & 2 & 5/2      \\
 $^1D_2$  & 2 & 0 & 2 & 4 & 7/2      \\
[1.1ex]
\hline\hline
\end{tabular}
\label{table4}
\end{center}
\end{table}

\subsection*{3.3 The state $1/2^-$}

The partial wave quantum numbers for that state are shown in Table 7. In this
 case there are only $5$ channels. It was necessary to enhance the $^3P_2$ and
 $^1D_2$ $nn$ forces to reach a $3n$ bound state. For $\lambda _{^3P_2}=3.5$
 and $\lambda_{^1D_2}=6.90$ the $2n$ bound states just disappear and the $3n$
 system is bound with $-1.90$ MeV. Again we display some intermediate
 resonance positions in Table 8. 

The final $S$-matrix pole position is significantly further away from the
positive real energy axis than for the states $3/2^{\pm}$. The resonance
trajectory is shown in Fig.\ 4. If one include $^3P_1$, one has $7$ channels
and the resonance position shifts by $-4.65-i~0.53$ MeV which is insignificant
in relation to the value without $^3P_1$.

This case poses even more difficulties to compare to Ref.\ [16], since there
 $^3P_0$ and $^3P_2-^3F_2$ have been enhanced while here we found it more
 natural to enhance  $^3P_2$ and $^1D_2$ (the latter force component was not
 included in Ref.\ [16]). For  the enhancements $\lambda _{^3P_0}=5.5$  and
 $\lambda _{^3P_2-^3F_2}=2.75$ in [16] the $3n$ resonance  turned out to be
 $2.82-i~2.82$ MeV. Apparently these factors are still quite large that also
 here one cannot expect a nearby resonance.

\begin{table}
\begin{center}
\caption{\small Intermediate $3n$ resonance positions  together with the
enhancement factor \( \lambda _{3_{P_{2}}} \) and  \( \lambda _{3_{P_{0}}}\)
  for a 10 channel calculation and $J^{\pi} = \frac{3}{2}^{+}$.} 
\vspace*{0.3cm}
\begin{tabular}{||c|c|c||}
\hline\hline
{$\lambda _{^3P_2}$}  &{$\lambda _{^3P_0}$}& $E^{res.}_{3n}$ [MeV]  \\
[1.1ex]
\hline\hline 
 3.20  & 4.70 &    0.60-$i$ 0.004    \\
 3.00  & 4.50 &    3.21-$i$ 0.96     \\
 2.00  & 3.50 &    6.41-$i$ 14.75     \\
 1.80  & 3.30 &    5.12-$i$ 18.18     \\
 1.00  & 2.50 &   -3.53-$i$ 29.96     \\
 1.00  & 1.00 &  -10.30-$i$ 33.20    \\
[1.1ex]
\hline\hline
\end{tabular}
\label{table5}
\end{center}
\end{table}

\begin{table}
\begin{center}
\caption{\small The final $3n$ resonance positions for the state
$J^{\pi} = \frac{3}{2}^{+}$ including different number of channels.}

\vspace*{0.3cm}
\begin{tabular}{||c|c||}
\hline\hline
{\rm Number of channels}  & {\rm Final resonance positions}  \\
[1.1ex]
\hline\hline 
 3  &  -9.82-$i$ 58.37    \\
 4  & -10.06-$i$ 58.31    \\
 5  & -10.24-$i$ 33.16     \\
 6  & -10.25-$i$ 33.17     \\
 7  & -10.61-$i$ 33.37     \\
 8  & -10.46-$i$ 33.29    \\
 9  & -10.35-$i$ 33.23     \\
 10 & -10.30-$i$ 33.20    \\
[1.1ex]
\hline\hline
\end{tabular}
\label{table6}
\end{center}
\end{table}

\begin{figure}
\centering
\psfig{file=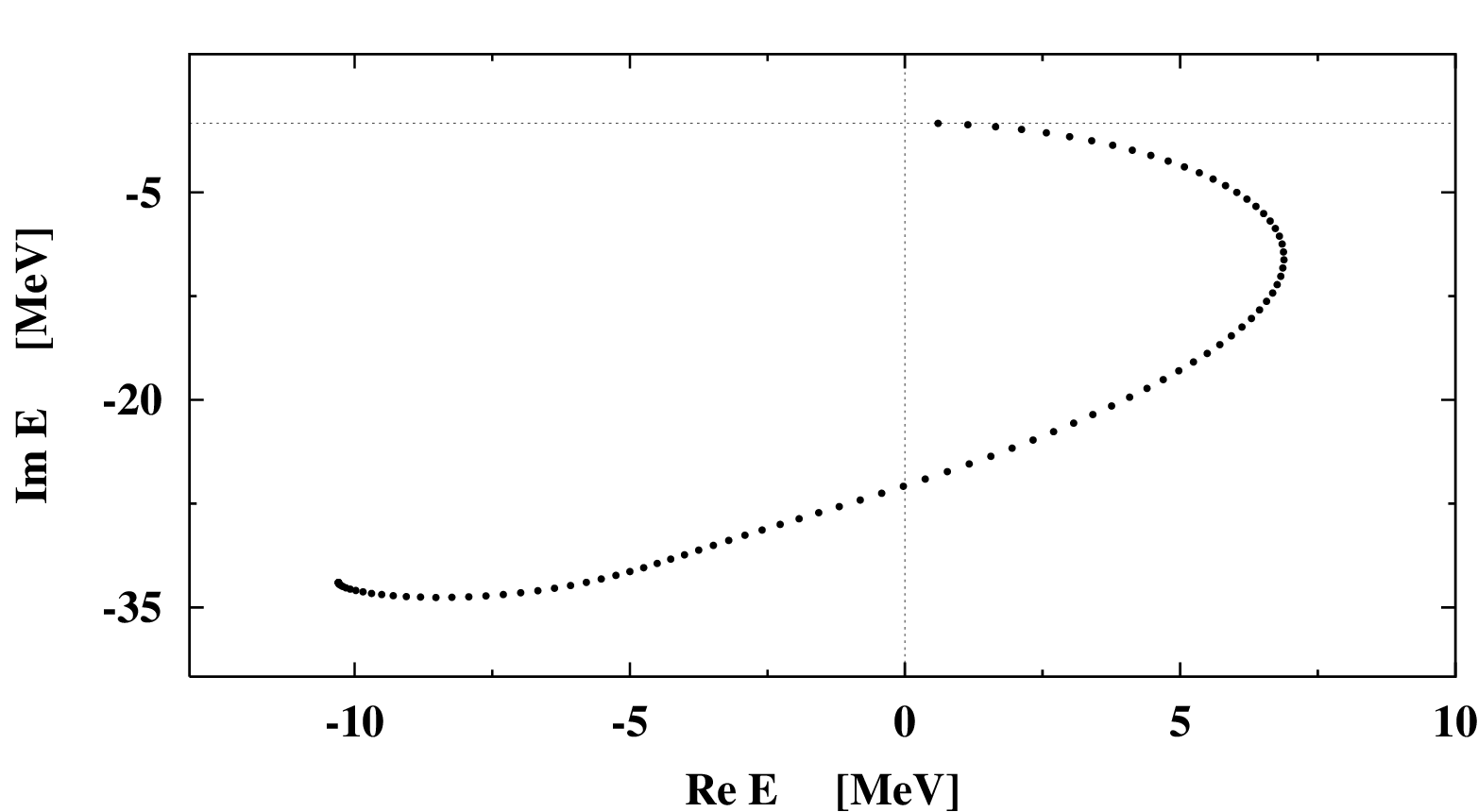,width=13cm}
\caption [Fig.2] {The resonance pole trajectory for the state $3/2^+$.} 
\end{figure}
\vspace{0.7cm}

\begin{table}
\begin{center}
\caption{\small The partial wave quantum number for the three neutron state
  $J^{\pi} = \frac{1}{2}^{-}$.}
\vspace*{0.3cm}
\begin{tabular}{||c|c|c|c|c|c||}
\hline\hline
 $nn$ {\rm state} & $l$ & $s$ & $j$ & $\lambda$ & $I$   \\
[1.1ex]
\hline\hline 
 $^1S_0$  & 0 & 0 & 0 & 1 & 1/2     \\
 $^3P_0$  & 1 & 1 & 0 & 0 & 1/2     \\
 $^3P_2$  & 1 & 1 & 2 & 2 & 3/2     \\
 $^3P_2$  & 1 & 1 & 2 & 2 & 5/2     \\
 $^1D_2$  & 2 & 0 & 2 & 1 & 3/2     \\
[1.1ex]
\hline\hline
\end{tabular}
\label{table7}
\end{center}
\end{table}

\begin{table}
\begin{center}
\caption{\small Intermediate $3n$ resonance positions  together with the
enhancement factor \( \lambda _{3_{P_{2}}} \) and  \( \lambda _{1_{D_{2}}}\)
  for a 5 channel calculation and $J^{\pi} = \frac{1}{2}^{-}$.} 
\vspace*{0.3cm}
\begin{tabular}{||c|c|c||}
\hline\hline
{$\lambda_{^3P_2}$}  &{$\lambda_{^1D_2}$}& 
 $E^{res.}_{3n}$ [MeV]  \\
[1.1ex]
\hline\hline 
 3.38  & 6.88 &    0.09-$i$ 0.0002    \\
 3.00  & 6.50 &    7.32-$i$ 3.49    \\
 2.00  & 5.50 &    16.44-$i$ 23.16     \\
 1.50  & 5.00 &    14.51-$i$ 32.15     \\
 1.00  & 4.00 &    6.10 -$i$ 42.27     \\
 1.00  & 3.00 &   -2.25 -$i$ 46.45    \\
 1.00  & 2.00 &   -11.36-$i$ 47.65    \\
 1.00  & 1.00 &   -20.38-$i$ 45.27    \\
[1.1ex]
\hline\hline
\end{tabular}
\label{table8}
\end{center}
\end{table}

\begin{figure}
\centering
\psfig{file=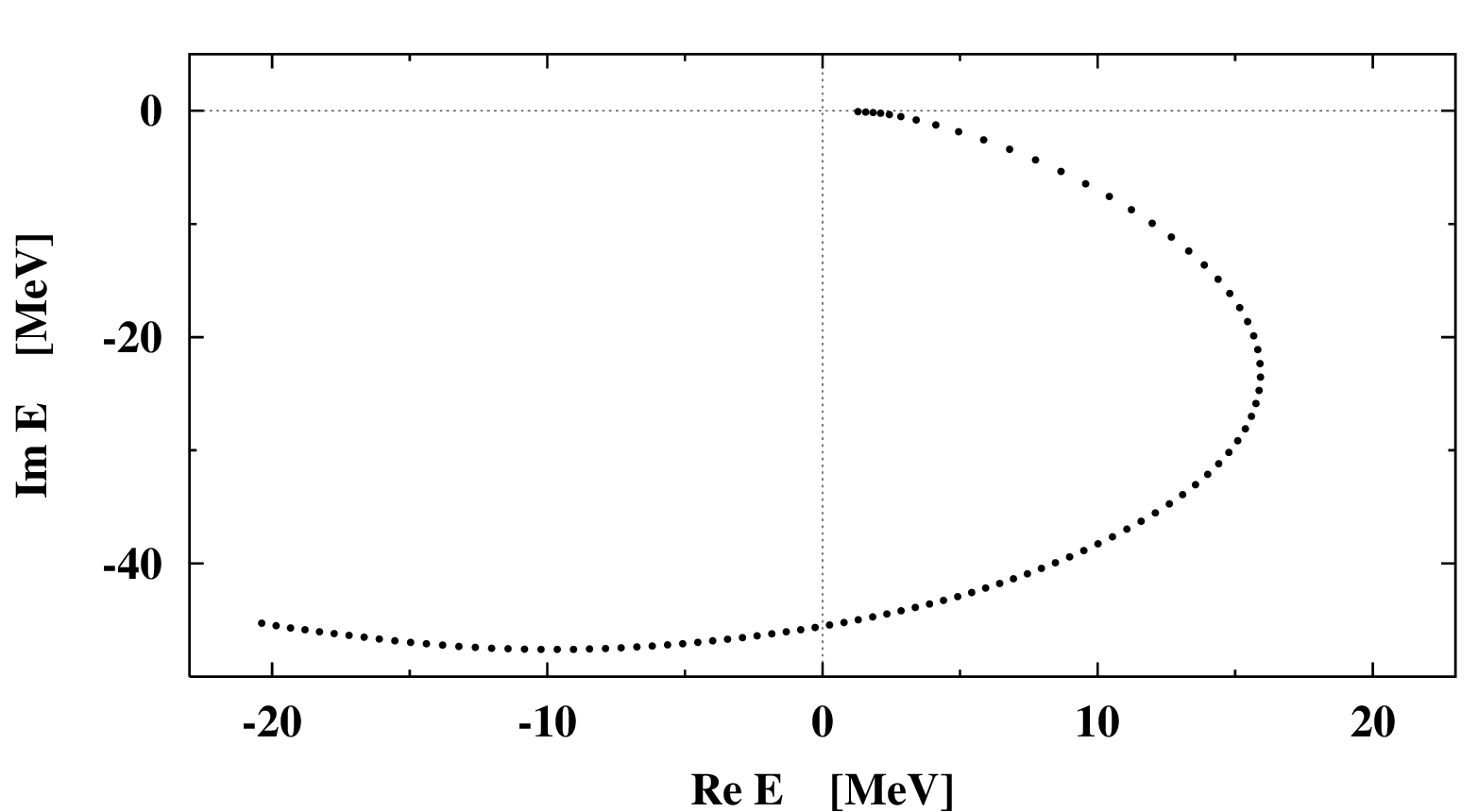,width=13cm}
\caption [Fig.2] {The resonance pole trajectory for the state $1/2^-$.} 
\end{figure}
\vspace{0.7cm}

\subsection*{3.4 The state $1/2^+$}

Here we used $6$ channels displayed in Table 9. Again we need to enhance both
the  $^3P_2$ and $^1D_2$ $nn$ forces by $3.5$ and $6.90$, respectively, to get
a  $3n$ bound state at $-7.38$ MeV. Some intermediate resonance positions are
shown in Table 10. The corresponding trajectory is shown in Fig.\ 5. The
inclusion of  $^3P_1$ leads to $8$ channels and makes a tiny shift to the
final resonance position of $-0.36-i~0.02$ MeV. In Ref.\ [16] this state has
not been considered. 
\begin{table}
\begin{center}
\caption{\small The partial wave quantum number for the three neutron state
  $J^{\pi} = \frac{1}{2}^{+}$.}
\vspace*{0.3cm}
\begin{tabular}{||c|c|c|c|c|c||}
\hline\hline
 $nn$ {\rm state} & $l$ & $s$ & $j$ & $\lambda$ & $I$   \\
[1.1ex]
\hline\hline 
 $^1S_0$  & 0 & 0 & 0 & 0 & 1/2     \\
 $^3P_0$  & 1 & 1 & 0 & 1 & 1/2    \\
 $^3P_2$  & 1 & 1 & 2 & 1 & 3/2     \\
 $^3P_2$  & 1 & 1 & 2 & 3 & 5/2     \\
 $^1D_2$  & 2 & 0 & 2 & 2 & 3/2     \\
 $^1D_2$  & 2 & 0 & 2 & 2 & 5/2     \\
[1.1ex]
\hline\hline
\end{tabular}
\label{table9}
\end{center}
\end{table}

\begin{table}
\begin{center}
\caption{\small Intermediate $3n$ resonance positions  together with the
enhancement factor $\lambda _{^3P_2}$ and $\lambda _{^1D_2}$  for a $6$
channel calculation and $J^{\pi} = \frac{1}{2}^{+}$.}  
\vspace*{0.3cm}
\begin{tabular}{||c|c|c||}
\hline\hline
{$\lambda _{^3P_2}$}  &{$\lambda _{^1D_2}$}& 
 $E^{res.}_{3n}$ [MeV]  \\
[1.1ex]
\hline\hline 
 3.43  & 6.43 &    0.64 -$i$ 0.002    \\
 3.00  & 6.00 &    12.90-$i$ 4.46     \\
 2.00  & 5.00 &    16.91-$i$ 22.63     \\
 1.00  & 4.00 &    11.94-$i$ 34.46     \\
 1.00  & 3.00 &    3.97 -$i$ 41.41     \\
 1.00  & 2.00 &   -4.55 -$i$ 45.51    \\
 1.00  & 1.00 &   -14.29-$i$ 48.34    \\
[1.1ex]
\hline\hline
\end{tabular}
\label{table10}
\end{center}
\end{table}

\begin{figure}
\centering
\psfig{file=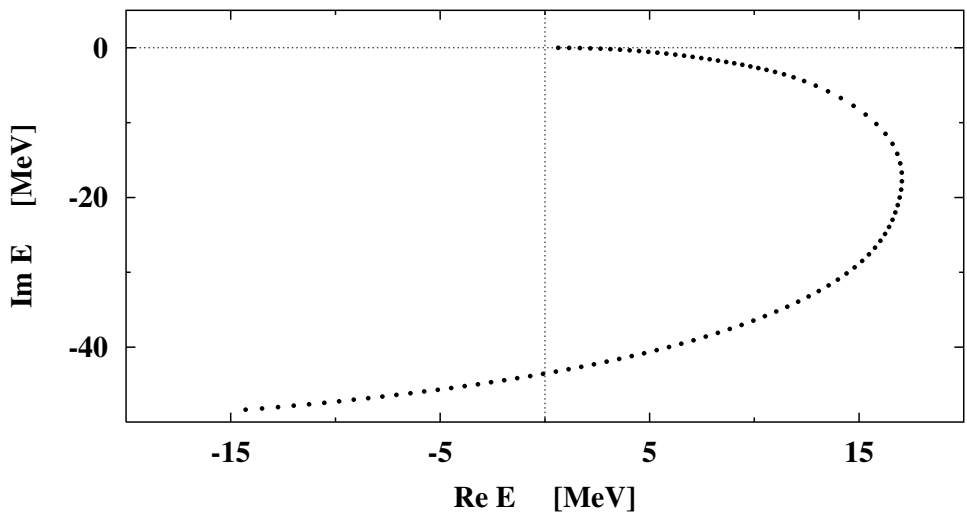,width=13cm}
\caption [Fig.5] {The resonance pole trajectory for the state $1/2^+$.} 
\end{figure}
\vspace{100.7cm}

\section*{IV Resonance Prediction via Pad\'e}

The interesting idea to predict low lying resonances with the help of
 Pad\'e approximants from a sequence of (auxiliary) bound state
 energies has been proposed in [22] and exemplified for instance in [23]. We
 also applied it to the system of $3n$'s. Assume, the $3n$ binding energy is
 given in terms of a power series in an enhancement factor $\lambda$ as
\begin {equation}
E(\lambda )=E({\lambda _{0}})+({\lambda -{\lambda _{0}}})^{\xi}a_{1}+
({\lambda -{\lambda _{0}}})^{\xi+1}a_{2}+\cdot\cdot\cdot\cdot\cdot\cdot
+({\lambda -{\lambda _{0}}})^{\xi+m-1}a_{m} 
\label{eq5}
\end{equation}
where $\xi$ has still to be determined.In searching for resonances close to 
the threshold $E=0$ it is natural to choose $\lambda _{0}$ such that
$E(\lambda_0)=0$. For two particles in this case it is known [22] that $\xi
=1$ for a $S$- wave and $\xi=\frac{1}{2}$ for $P$-waves and higher
ones. For a three-body system and the three-body break-up threshold we are
not aware of an analytical insight into the value $\xi$. We determined
$\xi$ numerically by approaching $E=0$ from both sides, $\lambda$ greater
than $\lambda_{0}$ and $\lambda$ smaller than $\lambda_{0}$. We found 
$\xi=\frac{1}{2}$ within our numerical accuracy for all studied cases 
$J^{\pi}=1/2^{\pm}$ and $J^{\pi}=3/2^{\pm}$. Choosing $m$ $3n$ bound state 
energies close to $E=0$ one can determine  $a_{1}$, $a_{2}$,
$\cdot\cdot\cdot$ $a_{m}$. To do that we solved the resulting $m$
equations by the method of singular value decomposition [24]. The power series
(6) can then  be  rewritten in form of Pad\'e approximants. We used a
continued fraction expansion, which is equivalent to certain orders of
approximants(see for instance [19]).  Choosing for instance $m=6$ one can reach the Pad\'e 
approximant of the order [3,3]. This turned out to be sufficient to predict the low lying
resonances  displayed  in Tables 12,14,16 and 18. In addition  we show the
bound state  energies in Tables 11,13,15 and 17 out of which the $a_{m}$
have been determined.  There we also include the resonance positions
determined directly from the analytically continued Faddeev equation. In the
cases $J^{\pi}={\frac{3}{2}}^+$ and ${\frac{1}{2}}^{\pm}$ we changed two
enhancement factors at the same time, but exactly by the same amount
$\Delta\lambda$. In this case  $\Delta\lambda$ is the expansion parameter in  
Eq.\ ({\ref{eq5}}). The $a_m$'s as well as the resonance energies are then
determined either through the enhancement factors in one or the other 
state. The resulting resonance positions are exactly the same. The last rows
in Tables $13$, $15$ and $17$ show the two  
 $\lambda_0$-values to many digits required to reach a significantly small
 $3n$ binding energy.
In all cases we see a very  good agreement of the low lying resonances using 
the Pad\'e method with the directly calculated ones. It is important
to note that the $3n$ bound state energies which serve as input have to be
known to $5$ digits in order to achieve the quoted Pad\'e results.


\begin{table}
\begin{center}
\caption{\small The $J^{\pi} = \frac{3}{2}^{-}$ bound state energies used as
  input for Pad\'e [3,3].  The last row shows $\lambda_0$ and the
  resulting energy.} 
\vspace*{0.3cm}
\begin{tabular}{||c|c|c||}
\hline\hline
{$ \lambda _{^3P_2}$}  & $E^{3n}_{B}$ [MeV]  \\
[1.1ex]
\hline\hline 
 3.41      & -4.79366    \\
 3.36      & -3.28446    \\
 3.31      & -1.89909     \\
 3.26      & -0.65666     \\
 3.25      & -0.42848     \\
 3.24      & -0.20848    \\
 3.2300873 & -1.1$\times 10^{-8}$ \\
[1.1ex]
\hline\hline
\end{tabular}
\label{table11}
\end{center}
\end{table}

\begin{table}
\begin{center}
\caption{\small The  $J^{\pi} = \frac{3}{2}^{-}$ $3n$ resonance positions
  determined  directly from the Faddeev equation in comparison to the Pad\'e [3,3] prediction.} 
\vspace*{0.3cm}
\begin{tabular}{||c|c|c||}
\hline\hline
{$ \lambda _{^3P_2}$}  & $E^{3n}_{res}$ [Me] via Faddeev
 & Pad\'e[3,3] [MeV] \\
[1.1ex]
\hline\hline 
 3.22      & 0.20-$i$ 0.003     & 0.20-$i$ 0.003  \\
 3.21      & 0.39-$i$ 0.014     & 0.39-$i$ 0.012 \\
 3.20      & 0.57-$i$ 0.03      & 0.57-$i$ 0.03   \\
 3.19      & 0.74-$i$ 0.06      & 0.74-$i$ 0.05  \\
 3.18      & 0.91-$i$  0.09      & 0.90-$i$ 0.08   \\
 3.17      & 1.08-$i$ 0.13      & 1.06-$i$ 0.12  \\
 3.16      & 1.24-$i$ 0.18      & 1.21-$i$ 0.16  \\
 3.15      & 1.40-$i$ 0.23      & 1.35-$i$ 0.22   \\
 3.14      & 1.56-$i$ 0.28      & 1.50-$i$ 0.28   \\
 3.13      & 1.71-$i$ 0.34      & 1.63-$i$ 0.35   \\
[1.1ex]
\hline\hline
\end{tabular}
\label{table12}
\end{center}
\end{table}
\vspace{.7cm}

\begin{table}
\begin{center}
\caption{\small The $J^{\pi} = \frac{3}{2}^{+}$ bound state energies used as
  input for Pad\'e [3,3]. For the last row see text.} 
\vspace*{0.3cm}
\begin{tabular}{||c|c|c||}
\hline\hline
{$\lambda _{^3P_2} $}  & {$ \lambda _{^3P_0}$}& $E^{3n}_{B}$ [MeV]  \\
[1.1ex]
\hline\hline 
 3.41         & 4.91         & -4.04286 \\
 3.36         & 4.86         & -2.76896 \\
 3.31         & 4.81         & -1.58830 \\
 3.26         & 4.76         & -0.51213 \\
 3.25         & 4.75         & -0.31107 \\
 3.24         & 4.74         & -0.11532 \\
 3.2339687436 & 4.7339687436 & -1.4$\times 10^{-9}$ \\
[1.1ex]
\hline\hline
\end{tabular}
\label{table13}
\end{center}
\end{table}
\vspace{0.2cm}

\begin{table}
\begin{center}
\caption{\small The  $J^{\pi} = \frac{3}{2}^{+}$ $3n$ resonance positions
 determined directly from the Faddeev equation in comparison to the Pad\'e [3,3] prediction.} 
\vspace*{0.3cm}
\begin{tabular}{||c|c|c|c||}
\hline\hline
{$ \lambda _{^3P_2} $}  &{$\lambda _{^3P_0}$} &  $E^{3n}_{res}$ [MeV]
 & Pad\'e [3,3] [MeV]  \\
[1.1ex]
\hline\hline 
 3.22  & 4.72    & 0.26-$i$ 0.0001    & 0.26-$i$ 0.0006  \\
 3.21  & 4.71    & 0.43-$i$ 0.001     & 0.43-$i$ 0.003  \\
 3.20  & 4.70    & 0.60-$i$ 0.004     & 0.61-$i$ 0.007   \\
 3.19  & 4.69    & 0.76-$i$ 0.012     & 0.77-$i$ 0.013   \\
 3.18  & 4.68    & 0.92-$i$ 0.024     & 0.93-$i$ 0.022   \\
 3.17  & 4.67    & 1.06-$i$ 0.04      & 1.08-$i$ 0.03   \\
 3.16  & 4.66    & 1.21-$i$ 0.06      & 1.23-$i$ 0.05   \\
 3.15  & 4.65    & 1.35-$i$ 0.09     & 1.37-$i$ 0.07    \\
 3.14  & 4.64    & 1.49-$i$ 0.13      & 1.50-$i$ 0.09    \\
 3.13  & 4.63    & 1.62-$i$ 0.17      & 1.63-$i$ 0.11    \\
 3.12  & 4.62    & 1.76-$i$ 0.21      & 1.76-$i$ 0.14    \\
[1.1ex]
\hline\hline
\end{tabular}
\label{table14}
\end{center}
\end{table}
\vspace{.7cm}

\begin{table}
\begin{center}
\caption{\small The $J^{\pi} = \frac{1}{2}^{-}$ bound state energies used as
  input for Pad\'e [3,3]. For the last row see text.} 
\vspace*{0.3cm}
\begin{tabular}{||c|c|c||}
\hline\hline
{$ \lambda _{^3P_2}$}  & {$\lambda _{^1D_2}$}& $E^{3n}_{B}$ [MeV]  \\
[1.1ex]
\hline\hline 
 3.49         & 6.89         & -1.58303 \\
 3.48         & 6.88         & -1.27353 \\
 3.47         & 6.87         & -0.96991 \\
 3.46         & 6.86         & -0.67262 \\
 3.45         & 6.85         & -0.38224 \\
 3.44         & 6.84         & -0.09983 \\
 3.436380542 & 6.836380542   & -1.8$\times 10^{-7}$ \\
[1.1ex]
\hline\hline
\end{tabular}
\label{table15}
\end{center}
\end{table}
\vspace*{0.2cm}

\begin{table}
\begin{center}
\caption{\small The $J^{\pi} = \frac{1}{2}^{-}$ $3n$ resonance positions
  determined  directly from the Faddeev equation in comparison to the Pad\'e [3,3] prediction.} 
\vspace*{0.3cm}
\begin{tabular}{||c|c|c|c||}
\hline\hline
{$\lambda _{^3P_2}$}  &{$\lambda _{^1D_2}$} &  $E^{3n}_{res}$ [MeV] via
Faddeev  & Pad\'e [3,3] [MeV]  \\
[1.1ex]
\hline\hline 
 3.43  & 6.83    & 0.17-$i$ 0.001    & 0.17-$i$ 0.001  \\
 3.42  & 6.82    & 0.43-$i$ 0.009    & 0.43-$i$ 0.01  \\
 3.41  & 6.81    & 0.69-$i$ 0.02     & 0.69-$i$ 0.03   \\
 3.40  & 6.80    & 0.94-$i$ 0.05     & 0.94-$i$ 0.06   \\
 3.39  & 6.79    & 1.19-$i$ 0.07     & 1.19-$i$ 0.09   \\
 3.38  & 6.78    & 1.43-$i$ 0.10     & 1.44-$i$ 0.13 \\
 3.37  & 6.77    & 1.67-$i$ 0.14     & 1.69-$i$ 0.18  \\
 3.36  & 6.76    & 1.91-$i$ 0.18     & 1.94-$i$ 0.23   \\
[1.1ex]
\hline\hline
\end{tabular}
\label{table16}
\end{center}
\end{table}
 \vspace{0.5cm}
\begin{table}
\begin{center}
\caption{\small The $J^{\pi} = \frac{1}{2}^{+}$ bound state energies used as
  input for Pad\'e [3,3]. For the last row see text.} 
\vspace*{0.3cm}
\begin{tabular}{||c|c|c||}
\hline\hline
{$\lambda _{^3P_2}$}  & {$\lambda _{^1D_2}$}& $E^{3n}_{B}$ [MeV]  \\
[1.1ex]
\hline\hline 
 3.50         & 6.90         & -7.37894 \\
 3.45         & 6.85         & -5.02695 \\
 3.41         & 6.81         & -3.22567 \\
 3.36         & 6.76         & -1.08684 \\
 3.35         & 6.75         & -0.67596 \\
 3.34         & 6.74         & -0.27141 \\
 3.333194193  & 6.733194193  & -2.0$\times 10^{-7}$  \\
[1.1ex]
\hline\hline
\end{tabular}
\label{table17}
\end{center}
\end{table}
\vspace{0.3cm}
\begin{table}
\vspace{1.2cm}
\begin{center}
\centerline{\parbox{14cm}{
\caption{\small The $J^{\pi} = \frac{1}{2}^{+}$ $3n$ resonance positions
  determined directly from Faddeev equation   in comparison to Pad\'e [3,3]
  prediction.}}}  
\vspace*{0.3cm}
\begin{tabular}{||c|c|c|c||}
\hline\hline
{$\lambda _{^3P_2}$}  &{$\lambda _{^1D_2}$} &  $E^{3n}_{res}$ [MeV] via Faddeev
 & Pad\'e [3,3] [MeV]  \\
[1.1ex]
\hline\hline 
 3.33  & 6.73    & 0.13-$i$ 0.0000    & 0.13-$i$ 0.0002  \\
 3.32  & 6.72    & 0.52-$i$ 0.0007    & 0.52-$i$ 0.003  \\
 3.31  & 6.71    & 0.90-$i$ 0.004     & 0.90-$i$ 0.007   \\
 3.29  & 6.70    & 1.27-$i$ 0.009     & 1.28-$i$ 0.011   \\
 3.28  & 6.69    & 1.64-$i$ 0.02      & 1.66-$i$ 0.01    \\
 3.27  & 6.68    & 2.00-$i$ 0.03      & 2.02-$i$ 0.02   \\
 3.26  & 6.67    & 2.35-$i$ 0.05      & 2.38-$i$ 0.02   \\
 3.25  & 6.66    & 2.70-$i$ 0.07      & 2.73-$i$ 0.03    \\
 3.35  & 6.65    & 3.04-$i$ 0.09      & 3.07-$i$ 0.03    \\
[1.1ex]
\hline\hline
\end{tabular}
\label{table18}
\end{center}
\end{table}
\vspace{30.0cm}

\section*{V Summary}

We analytically continued the Faddeev equation for three neutrons into the
unphysical sheet of the complex energy plane adjacent to the positive real
energy axis. The nn forces are of finite rank and were kept in
the partial waves $^1S_0$, $^3P_0$, $^3P_1$, $^3P_2$, $^1D_2$. This leads up
to $13$ channel calculations.
The $^3P_1$ $nn$ force is repulsive and its effect turned out to be rather small in
relation to the combined effects of all the other forces. 
 The analytical continuation  is performed via 
a contour deformation of the path of integration. The strategy is to
artificially enhance the nn forces such that a  3n bound state is  generated. Then by reducing that enhancement one follows the
energy eigenvalue of the homogeneous 3n Faddeev equation which
traces out a trajectory. After  having reached the 3n break-up
threshold $E=0$ that trajectory dives into the  unphysical energy sheet. The
final positions for the physical strength turned out for the four cases
studied, $J^{\pi}=1/2^{\pm}$ and $J^{\pi}=3/2^{\pm}$, to be far away from the 
positive real energy axis. Though the nn forces employed are 
not of the quality of the modern high precision $NN$ potentials it appears
unlikely that those modern forces would lead to totally different
results. Thus we conjecture that 3n resonances close to the
physical region will not exist. This conjecture is also supported by the
findings in \ [16].

These results gained through the analytically
continued Faddeev equation  have been tested numerically using Pad\'e
approximants. In this method one generates a set of auxiliary 3n bound state energies close to $E=0$ for different enhancement factors $\lambda$.
This determines the first few coefficients in a power series expansion of
 $E(\lambda)$. This  again leads to the coefficients of Pad\'e
 approximants which can be used to predict low  lying 3n  resonances. The
 resulting values agreed very well  with the ones determined
 via the Faddeev equation.

\begin{table}
\begin{center}
\caption{The parameters of our $nn$ potential. The units of these parameters 
are fm$^{-1}$ for $\beta _{ij}$,
MeV fm$^{-2l-1}$ for $\lambda _{i}$ and the $\gamma _{i}$ are dimensionless.}
\vspace{0.3cm}
\begin{tabular}{||c|c|c|c|c|c||}
\hline\hline
                     & $^1S_0$ & $^3P_0$ & $^3P_1$ & $^3P_2$ &  $^1D_2$ \\
[1.1ex]
\hline\hline 
\( \beta  \)\( _{11} \) & 0.8131678&0.8322894& 0.9713441& 1.977127 & 2.522169  \\
\( \beta  \)\( _{12} \) & 1.288463 &1.262785 & 2.180297 & 3.147108 & 1.651999  \\
\( \beta  \)\( _{21} \) & 7.496476 &2.645783 &  &       &           \\
\( \beta  \)\( _{22} \) & 1.661389 &         &  &       &           \\
\( \gamma  \)\( _{1} \) & 2.698168 &6.397468 & 31.97981 &5.242138 & -0.2668325\\
\( \gamma  \)\( _{2} \) & 0.3270664&         &  &        &           \\
\( \lambda  \)\( _{1} \)& -17.87098&-12.10082& 59.89249  &  -1477.246 &-918408.7 \\
\( \lambda  \)\( _{2} \)& 82710.93 &277071.8 &  &  &           \\
[1.1ex]
\hline\hline
\end{tabular}
\end{center}
\end{table}

\begin{table}
\begin{center}
\caption{The $nn$ effective range parameter $r$ and the scattering length $a$
                     in units of fm for the  $^1S_0$-state in comparison with
                     experimental results.}
\vspace{0.3cm}
\begin{tabular}{||c|c|c||}
\hline\hline
    & $r$ & $a$  \\
[1.1ex]
\hline\hline 
our potential       &   2.85 & -19.08  \\
exp. data [21]   & 2.83\( \pm  \)0.11 &-18.6\( \pm  \)0.5  \\
[1.1ex]
\hline\hline
\end{tabular}
\end{center}
\end{table}

\section*{Acknowledgements}
\setcounter{equation}{0}
We would like to thank Y.\ Koike for very helpful hints at the beginning of
 the study. A.\ Hemmdan would like to thank the Institut f\"ur Theoretische
 Physik II of the Ruhr-Universit\"at Bochum for the very kind
 hospitality. This research was supported by the Egyptian government under a
 fellowship in the Channel System.

\section*{Appendix A}
The finite rank potential we use has  the following form~[25] 

\begin{equation} 
V(p,p\prime)=\sum^{2}_{i=1} g_i(p)\lambda_{i}g_i(p\prime)
\end{equation}
where 
\begin{equation}
g_1(p)=\frac{p^l}{(p^2+\beta^2_{11})^{l+1}}+\frac{p^{l+2}}{(p^2+\beta^2_{12})^{l+2}}\gamma_{1}
\end{equation}
\begin{equation}
g_2(p)=\frac{p^{l+2}}{(p^2+\beta^2_{21})^{l+2}}+\frac{p^{l+4}}{(p^2+\beta^2_{22})^{l+3}} \gamma_{2}
\end{equation}

\vspace{1.cm}
The $l$-dependent parameters of the potential form factors are displayed in 
Table 19.  The resulting phase shifts of the nn system for the 
 partial waves  $^1S_0$, $^3P_0$, $^3P_1$, $^3P_2$ and $^1D_2$ are compared to $pp$
 phase shift values (without Coulomb) given in SAID [21] in Fig.\ 6. In Table 20 we
 show the scattering length $a$ and the effective range $r$ for the $nn$
 system in the state  $^1S_0$.

\vspace{0.7cm}
\begin{figure}
\centering
\psfig{file=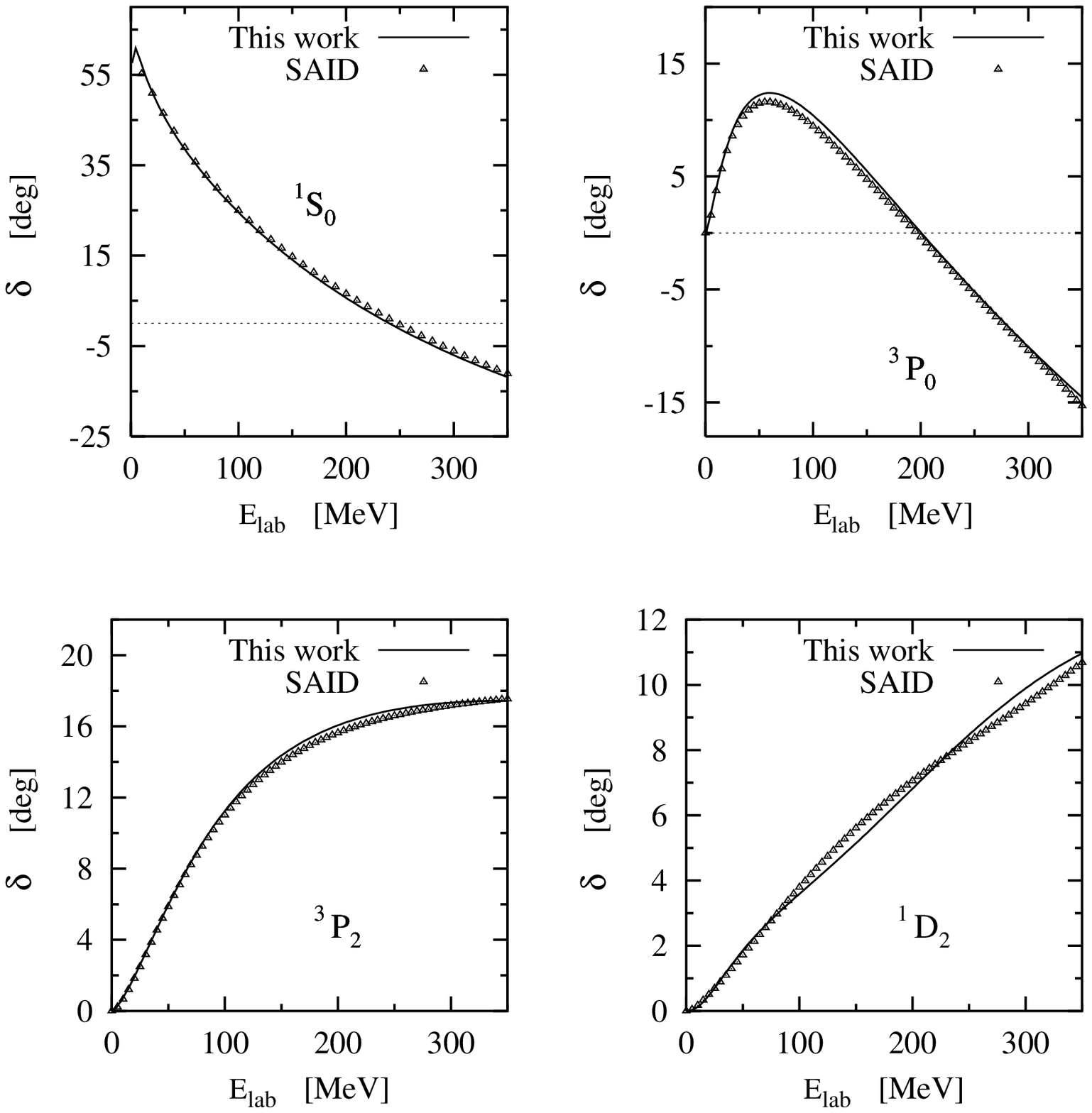,width=15cm}
\psfig{file=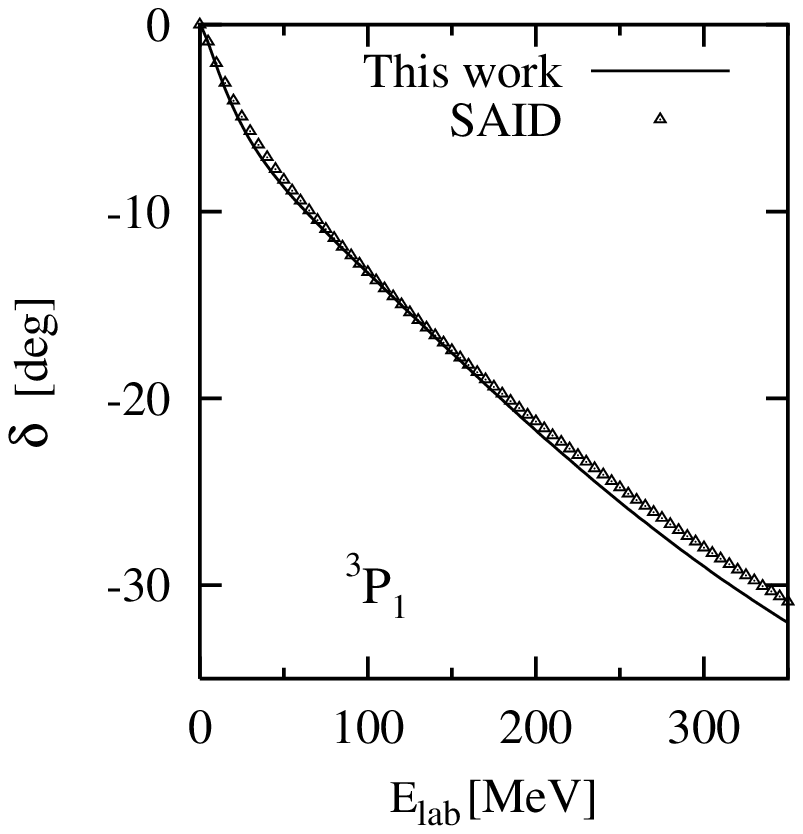,width=7cm}
\caption [Fig.52] {The calculated $nn$ phase shifts  compared to SAID [21] $pp$
phase shift values (without Coulomb).} 

\end{figure}
\vspace{0.7cm}

\vspace{100.5cm}
\newcommand{\etal}{\it et al.}
\setlength{\parindent}{0mm}

\vspace{500cm}
{\bf References}
\begin{list}{}{\setlength{\topsep}{0mm}\setlength{\itemsep}{0mm}%
\setlength{\parsep}{0mm}}
%
\item[1.]  D.\ R.\ Tilley, H.\ R.\ Weller and H.\ H.\ Hasan, Nucl.\ Phys.\
  {\bf A474}, 1 (1987). 
\item[2.]  M.\ Yuly {\it et al.}, Phys.\ Rev.\ {\bf C55}, 1848 (1997).
\item[3.]  J.\ Sperinde, D.\ Frederickson, R.\ Hinkins, V.\ Prez-Mendez and
  B.\ Smith, Phys.\ Lett.\ {\bf 32B}, 185 (1970).
\item[4.]  A.\ Setz {\it et al.}, Nucl.\ Phys.\ {\bf A457}, 669 (1986).
\item[5.]  F.\ M.\ Marques {\it et al.}, Phys.\ Rev.\ {\bf C65}, 044006 (2002).
\item[6.]  S.\ A.\ Coon and H.\ K.\ Han, Few-Body System {\bf 30}, 131 (2001).
\item[7.]  B.\ S.\ Pudliner {\it et al.}, Phys.\ Rev.\ {\bf C56}, 1720 (1997).
\item[8.]   W.\ Gl\"ockle. Phys.\ Rev.\ {\bf C18}, 564 (1978).
\item[9.]   K.\ M\"oller and Yu.\ Orlov, Sov.\ J.\ Part.\ Nucl. {\bf 20}, 569
  (1989). 
\item[10.]  R.\ Offermann and W.\ Gl\"ockle, Nucl.\ Phys.\ {\bf A318}, 138
  (1979). 
\item[11.]  J.\ R.\ Reid, Ann.\ Phys. {\bf 50}, 411 (1968).
\item[12.]  S.\ A.\ Sofianos, S.\ A.\ Rakityansky and G.\ P.\ Vermaark, J.\
  Phys.\ G: Nucl.\ Part.\ Phys. {\bf 23}, 1619 (1997).
\item[13.]  A.\ Csoto, H.\ Oberhummer and R.\ Pichler, Phys.\ Rev.\ {\bf C53},
  1589 (1986). 
\item[14.]  D.\ R.\ Thompson, M.\ LeMere and Y.\ C.\ Tang, Nucl.\ Phys.\ {\bf
    A268},  53 (1977); 
  I.\ Reichstein and Y.C.Tang, Nucl.\ Phys.\ {\bf A158}, 529 (1970);
            A.\ Csoto, Phys.\ Rev.\ {\bf C48}, 165 (1993).
\item[15.]  P.\ Heiss and H.\ H.\ Hackenbroich, Phys.\ Lett.\ {\bf 30B}, 373
  (1969); 
  A.\ Csoto, R.\ G.\ Lovas and A.\ T.\ Kruppa, Phys.\ Lett.\ {\bf 70}, 1389
  (1993). 
\item[16.]  H.\ Wita\l a and W.\ Gl\"ockle, Phys.\ Rev.\  {\bf C60}, 024002 (1999).
\item[17.]  D.\ Gogny, P.\ Pires and R.\ de Tourreil, Phys.\ Lett.\ {\bf
    32B}, 59, (1970).  
\item[18.]  V.\ G.\ Stoks {\it et al.}, Phys.\ Rev.\ {\bf C49}, 2950 (1994).
\item[19.]  W.\ Gl\"ockle, {\it The Quantum Mechanical Few-Body Problem},
  Springer  Verlag, Heidelberg, 1983.
\item[20.]  A.\ Matsuyama and K.\ Yazaki, Nucl.\ Phys.\ {\bf A534}, 620 (1991).
\item[21.] http://NN-online.sci.kun.nl/index.html.
\item[22.] V.\ I.\ Kukulin {\it et al.}, Sov.\ J.\ Nucl.\ Phys.\ {\bf 92}, 421 (1979).
\item[23.] N.\ Tanaka, Y.\ Suzuki, K.\ Varga and R.\ G.\ Lovas, Phys.\ Rev.\
  {\bf C59}, 1391 (1999).
\item[24.] Numerical Recipes, Cambridge University Press (1992).
\item[25.] W. Schweiger, W. Plessas, L. P. Kok, H. van Haeringen, Phys. Rev. C {\bf 27},
515 (1983). 
%
\end{list}

\end{document}